Editorial Manager(tm) for Astrophysics and Space Science
Manuscript Draft

Manuscript Number: ASTR2861R1

Title: X-Ray Spectral Study of AGN Sources Content in Some Deep Extragalactic XMM-Newton  Fields

Article Type: Original research

Corresponding Author: Badie Abdel-halim Korany

Corresponding Author's Institution:

First Author: M. A. Hassan

Order of Authors: M. A. Hassan;Badie Abdel-halim Korany;R. Misra;I. A. M. Issa;M. K. Ahmed;F. A. Abdel-Salam

Abstract: We undertake a spectral study of a sample
of bright X-ray sources  taken from six XMM-Newton fields at high galactic latitudes,where AGN are the most populous class. These six fields were chosen such that the observation had an exposure time more than 60 ksec, had data from the EPIC-pn detector in the full-Frame mode and lying at high galactic latitude $|b| > 25^o$ .
The analysis started by fitting the spectra of all sources with an absorbed power-law model, and then we fitted all the spectra with an absorbed power-law with a low energy black-body component model.The sources for which we added a black body gave an F-test probability of 0.01 or less (i.e. at 99\% confidence level), were recognized as sources that display soft excess. We perform a comparative
analysis of soft excess spectral parameters with respect to the underlying power-law one for sources that satisfy this criterion. Those sources, that do not show evidence
for a soft excess, based on the F-test probability at a 99\%
confidence level, were also fitted with the absorbed power-law with a low energy black-body component model with the black-body temperature fixed at 0.1 and 0.2 keV.  We establish upper limits on the soft excess flux for those sources at these two temperatures. Finally we have made use of Aladdin interactive sky atlas and matching with NASA/IPAC Extragalactic Database (NED) to identify the X-ray sources in our sample. For those sources which
are identified in the NED catalogue, we make a comparative study of the soft excess phenomenon for different types of systems.

Response to Reviewers: Editors comments:

Dear Authors,
Kindly check the references and ensure that they are used in the text of your paper. At a glance, only about 8 appear to be referred to.

corrected

# X-Ray Spectral Study of AGN Sources Content in Some Deep Extragalactic XMM-Newton Fields

M. A. Hassan[1]. B. A. Korany[2]. R. Misra[3]. I. A. M. Issa[1]. M. K. Ahmed[4]. And F. A. Abdel-Salam[5]

1-National Research Institute of Astronomy and Geophysics, Helwan, Cairo, Egypt
2-The Faculty of Applied Science, Umm AL-Qura University, Makka, Saudi Arabia
3- Inter-University Center for Astronomy and astrophysics (IUCAA), India
4- Faculty of Science, Cairo University, Egypt
5- Faculty of Science, Taibah University, Madina, Saudi Arabia

*Abstract*

We undertake a spectral study of a sample of bright X-ray sources taken from six XMM-Newton fields at high galactic latitudes, where AGN are the most populous class. These six fields were chosen such that the observation had an exposure time more than 60 ksec, had data from the EPIC-pn detector in the full-Frame mode and lying at high galactic latitude $|b| > 25^o$. The analysis started by fitting the spectra of all sources with an absorbed power-law model, and then we fitted all the spectra with an absorbed power-law with a low energy black-body component model. The sources for which we added a black body gave an F-test probability of 0.01 or less (i.e. at 99% confidence level), were recognized as sources that display soft excess. We perform a comparative analysis of soft excess spectral parameters with respect to the underlying power-law one for sources that satisfy this criterion. Those sources, that do not show evidence for a soft excess, based on the F-test probability at a 99% confidence level, were also fitted with the absorbed power-law with a low energy black-body component model with the black-body temperature fixed at 0.1 and 0.2 keV. We establish upper limits on the soft excess flux for those sources at these two temperatures. Finally we have made use of Aladdin interactive sky atlas and matching with NASA/IPAC Extragalactic Database (NED) to identify the X-ray sources in our sample. For those sources which are identified in the NED catalogue, we make a comparative study of the soft excess phenomenon for different types of systems.

Keywords (X-rays: Galaxies: Active Galactic Nuclei)

# 1 Introduction

With the present generation of X-ray orbiting observatories, such as XMM-Newton and *Chandra*, our knowledge of the X-ray properties of Active Galactic Nuclei (AGN) has been enhanced. Long duration observations and detailed spectroscopy have provided new insights into the mechanisms of X-ray production, its modification by intervening matter, and X-ray variability. AGN emit a significant amount of radiation in the UV, which is commonly known as the ``big blue bump'' and can be modeled as thermal emission. They are also bright sources in X-rays, with a spectra which can be approximately represented as a power-law. AGN are thought to be powered by super-massive black holes accreting matter in the form of a disk. This accretion disk produces the UV emission observed in these sources, while active regions above the disk are responsible for the X-ray emission.

The X-ray emission from AGNs is characterized by an underlying power-law with photon index $\Gamma$ : 1.5 - 2 (Mushotzky et. al. 1993). Also, the X-ray spectra of AGNs have long been known to often contain a `soft excess` component at energies below : 1 keV (Turner & Pounds 1988). The soft excess is usually fitted with a black-body which has a roughly constant temperature of 0.1 - 0.2 keV over several decades in AGNs mass (Walter & Fink 1993, Czerny et. al. 2003, Gierlin'ski & Done 2004). AGNs with the largest soft excess are generally narrow-line Seyfert-1 galaxies (NLS1; Boroson 2002).

# 2 Observation s and data analysis

We are concentrating on som extragalactic sources(in particular AGNs), according to:a- Sky positions at high galactic latitudes $|b|> 25^o$, where AGNs are the most common X-ray sources .b- Observations for which we have data from the EPIC-pn(Struder et al., 2001) detector in Full-Frame mode (i.e. the EPIC-pn full FOV covered). c-Observation for which the total GTI is more than 60 Ksec. The geometrical shadowing of half of X-ray photons received by the MOS1 and MOS2(Turner et al. 2001) detectors makes the EPIC-pn camera a factor of two more sensitive than the MOS cameras. Therefore, we used only EPIC-pn exposures.

Six of the X-ray fields, available on XMM-Newton archive, which satisfied the above criteria are listed in Table (1) together with their observational details.

Table 1: Observational details of the XMM-Newton fields.

| Field | R.A. (j2000) | Dec. (j2000) | $b^1$ (deg) | $N_H^{Gal}$ ($10^{20} cm^{-2}$) | Rev./Obs.id | Obs. date | Filter[2] pn | GTI[3] pn |
|---|---|---|---|---|---|---|---|---|
| APM 08279+5255 | 08 31 41.57 | 52 45 17.7 | 36.24 | 3.83 | 0437/0092800201 | 2002-04-28 | M | 71.1 |
| COSMOS FIELD 21 | 09 58 26.40 | 02 42 36.0 | 41.55 | 2.76 | 0916/0203362101 | 2004-12-09 | Th | 60.8 |
| PG 1115+080 | 11 18 16.90 | 07 45 59.4 | 60.64 | 3.56 | 0825/0203560201 | 2004-06-10 | Th | 64.5 |
| POX52 | 12 02 56.90 | -20 56 03.0 | 40.55 | 4.21 | 1022/0302420101 | 2005-07-08 | M | 78.9 |
| PG 2112+059 | 21 14 52.60 | 06 07 42.0 | -28.55 | 6.61 | 1090/0300310201 | 2005-11-20 | Th | 73.1 |
| LBQS 2212-1759 | 22 15 31.67 | -17 44 05.7 | -52.93 | 2.39 | 0356/0106660601 | 2001-11-17 | Th | 89.9 |

1-Galactic latitude. 2-Blocking filter:Th:Thin;M:Medium. 3- Exposure time (in ks)per observation obtained after removal of background flares.

The SAS task *epchain* was used, with *ereject* turned on (5$\sigma$ and the default noise parameters), to generate the EPIC-pn calibrated photon event files. Subsequently, events with PHA values below 20 *adu* was removed using the SAS task *evselect*. The event files of the individual observations have been cleaned from the high flaring background periods. Only one field found to be free of flares . The SAS task *edetect_chain* was used to search for sources in the above energy bands with likelihood threshold (likemin= 12). From the list of sources detected in each field, we have taken the brightest ones (sources with at least 500 net counts in the 0.3 - 10 keV), only four bright sources have been rejected because they are diffuse sources. To extract the spectra we have defined a circular region of each object for each observation with a radius that varied depending on the position of the source within the detector, typically 14" - 30". The background regions were chosen to be located close to the source, with the same radius, and free from any contamination source. we have done the spectral analysis using XSPEC.

We start the analysis by fitting the spectra of all sources with an absorbed power-law model (hereafter APL model). Then we fitted all the spectra with an absorbed power-law with a low energy black-body component model (hereafter APLBB model). For both models, the absorption due to our Galaxy ($N_H^{Gal}$) of each field as is obtained by Nh command of Ftools. To identify the X-ray sources in our sample we have made use of the Aladin interactive sky atlas. Matching with NASA/IPAC Extragalactic Database (NED) led us to identify 26 of the sources as AGN or QSO with one of them being a lensed QSO. 17 of the sources are listed in the catalogue as either unidentified X-ray , Radio, Visual or UV excess sources, while the rest 38 were not found in the database. The digitized images of the six fields overlayed by the NED catalogue and the sources of the sample are shown in Appendix A. Those sources, for which adding a black-body gave a F-test probability of 0.01 or less (i.e. at a 99% confidence level), were recognized as sources that display soft excess. It was found that : 17% (14 out of 81 sources) of our sample satisfied this criterion. The NED catalogue identifies six of them as Quasars and AGN with redshifts ranging from $0.022$ to $3.9$.

Four of the other eight source are marked as unidentified X-ray sources in NED, while the rest four are not found in the database. Table (0) gives the details of the spectral fitting, identification with the NED catalogue, redshifts, and the intrinsic luminosities of those six sources identified as Quasars or AGN. The first source in this table have $\Gamma = -0.5$ and low luminosity ($= 8.575E41 erg/sec$) , its most probably from the Compton-thick source dominated by reflection. In this case the thermal emission may be due to the surrounding thermal plasma. The last two sources with $\Gamma$ about 0.4 look like partially ionized absorber, and in this case the 'soft excess' may be partially an artifact of the fitting the complex shape.The errors (the standard deviation) of $\Gamma$ is significantly larg. The effect of hte uncertainties of the x-ray power low index on the determination of the excesses are always larger than the typical uncertainties associated with the choice of the spectral model. This ensures that the excesses drived from the data are not affected by large systematic uncertainties linked with the choice of the model (Walter & Fink 1993). Table (1) gives the details of the spectral fitting and identification with NED for the rest ( eight sources ) that didn't identified as either Quasars or AGN.

Table  2: The six X-ray sources that have soft excess and identified as Quasars or AGN. Columms 1

and 2: Position of the X-ray source obtained by *edetect_chain* , column 3: APLBB model, column 6: Power-law photon Index, column 7: Fraction of black-body flux to total flux, column 8: Separation of X-ray source position from the NED position in arcsec. , column 9: Type of the best matched sources with our sample from NED, cloumn 10:The Redshift, cloumn 11: Intrinsic 0.3-10 Kev luminocity.

| $RA_X$ (j2000) | $Dec_X$ (j2000) | $\chi^2$ / dof | $N_H^{abs.}$ $10^{22} cm^{-2}$ | kT (keV) | $\Gamma$ | frac. | separ. (arcsec) | Type | R | Lumin (erg/sec) |
|---|---|---|---|---|---|---|---|---|---|---|
| 08:31:39.320 | +52:42:05.55 | 50.2/24 | $0.04^{+0.19}_{-0.04}$ | $0.16^{+0.05}_{-0.06}$ | $-0.5^{+0.2}_{-0.3}$ | 0.03 | 2.47 | G | 0.059 | 8.575E41 |
| 08:31:41.849 | +52:45:17.78 | 350.8/350 | $0.24^{+0.01}_{-0.02}$ | $0.19^{+0.03}_{-0.02}$ | $1.80^{+0.04}_{-0.11}$ | 0.20 | 2.39 | Q_Lens | 3.911 | 1.086E47 |
| 09:58:21.638 | +02:46:28.97 | 71.6/71 | $0.15^{+0.04}_{-0.03}$ | $0.10^{+0.02}_{-0.02}$ | $1.9^{+0.2}_{-0.2}$ | 0.30 | 1.12 | QSO | 1.403 | 1.839E45 |
| 11:18:17.066 | +07:45:58.20 | 373.6/394 | $0.12^{+0.02}_{-0.03}$ | $0.21^{+0.07}_{-0.06}$ | $1.84^{+0.10}_{-0.07}$ | 0.07 | 1.00 | G | 0.310 | 2.257E44 |
| 12:02:57.023 | -20:56:02.84 | 267.0/152 | $0.21^{+0.04}_{-0.02}$ | $0.080^{+0.004}_{-0.004}$ | $0.40^{+0.09}_{-0.07}$ | 0.34 | 1.73 | G | 0.022 | 5.331E41 |
| 21:14:52.510 | +06:07:42.51 | 69.1/46 | $0.22^{+0.20}_{-0.04}$ | $0.11^{+0.05}_{-0.02}$ | $0.5^{+0.2}_{-0.2}$ | 0.22 | 1.44 | QSO | 0.466 | 8.498E43 |

Table 3: The X-ray sources that have soft excess but not identified as either Quasars or AGN. Columns 1 and 2: Position of the X-ray source obtained by *edetect_chain* , column 3: APLBB model, column 6: Power-law photon index , column 7: Fraction of black-body flux to total flux, column 8: Separation of X-ray source position from the NED position in arcsec. , column 9: Type of the best matched sources with our sample from NED

| $RA_X$ (j2000) | $Dec_X$ (j2000) | $\chi^2$ / dof | $N_H^{abs.}$ $10^{22} cm^{-2}$ | kT (keV) | $\Gamma$ | frac. | separ. (arcsec.) | Type |
|---|---|---|---|---|---|---|---|---|
| 11:18:04.428 | +07:47:19.28 | 34.5 / 28 | $0.3^{+0.1}_{-0.2}$ | $0.09^{+0.02}_{-0.02}$ | $2.3^{+0.5}_{-0.4}$ | 0.70 | 13.76 | XrayS |
| 11:18:36.914 | +07:39:52.01 | 14.8 / 32 | $0.18^{+0.07}_{-0.04}$ | $0.09^{+0.02}_{-0.02}$ | $1.8^{+0.4}_{-0.3}$ | 0.50 | | |
| 12:02:59.353 | -20:52:36.69 | 187.5 / 104 | $0.21^{+0.03}_{-0.03}$ | $0.14^{+0.01}_{-0.02}$ | $0.9^{+0.8}_{-0.8}$ | 0.88 | | |
| 12:03:46.721 | -20:53:51.34 | 148.8 / 159 | $0.12^{+0.03}_{-0.04}$ | $0.07^{+0.02}_{-0.01}$ | $2.1^{+0.2}_{-0.1}$ | 0.23 | | |
| 21:14:26.097 | +05:57:23.64 | 50.1 / 32 | $0.07^{+0.06}_{-0.07}$ | $0.21^{+0.03}_{-0.05}$ | $0.6^{+0.3}_{-0.4}$ | 0.15 | 9.39 | XrayS |
| 21:14:48.955 | +06:18:57.99 | 132.0 / 103 | $0.27^{+0.14}_{-0.03}$ | $0.14^{+0.02}_{-0.03}$ | $2.5^{+0.6}_{-0.4}$ | 0.62 | 13.02 | XrayS |
| 21:15:13.728 | +06:06:25.73 | 59.1 / 39 | $0.7^{+0.3}_{-0.3}$ | $0.10^{+0.02}_{-0.02}$ | $2.2^{+1.0}_{-0.9}$ | 0.96 | | |
| 21:15:16.507 | +06:08:41.18 | 305.4 / 297 | $0.201^{+0.008}_{-0.024}$ | $0.078^{+0.010}_{-0.006}$ | $1.93^{+0.06}_{-0.10}$ | 0.31 | 6.38 | XrayS |

Figure (1) shows the histogram of the reduced $\chi^2$ for sources that exhibit soft excess. While for most of the sources the reduced $\chi^2$ :1, there are three notable exceptions where the reduced $\chi^2 > 1.5$, suggestive of other components or that for these sources the model used is not adequate. Figures (2) and (3) show the distribution of the soft excess temperature and power-law indicies for these sources. The soft excess temperatures are clustered around 0.1 keV, while the photon index varies from $-0.5$ to 2.5. Of these 14 sources, the NED catalogue identifies six of them as Quasars or AGN with redshifts ranging from $0.022$ to $3.9$ which implies that the intrinsic red-shift corrected temperature has a wider range $0.1-1.0$ keV. We have examined the results for possible correlations between the soft excess temperature, power-law photon index, $\Gamma$, and absorption column density $N_H^{abs.}$ by plotting them against each other (Figures 4 - 6). Visual inspection reveals that there are weak correlations: the temperature decreases with increasing $\Gamma$, as well as, with $N_H^{abs.}$. $N_H^{abs.}$ is correlated with $\Gamma$ which is suggestive that these correlations may be artifacts of low count statistics. This is because steep power-laws with large absorption resemble hard power-law spectra in these low energy bands. Moreover, statistical analysis of these correlation reveal that the results are dominated by a few bright sources biasing the sample. Hence we conclude that the statistics of this sample is not good enough to draw definite conclusions regarding any correlations between the spectral parameters.

We have studied the contribution of black-body flux to the total flux and we have found that the black-body flux represents more than 10% and 20% of the total flux for 85% and 79% of the sources respectively. Figure (7) shows the accumulative histogram of such a contribution. The fraction of the soft excess flux to the total flux has a surprisingly wide range ($0.03-0.96$). Of particular interest are three sources not identified by NED, that have extreme soft spectra with the soft excess dominating the luminosity at a fraction of $0.7$, $0.88$ and $0.96$ respectively. These soft sources have high inferred column density $>2\times10^{21}\ cm^{-2}$.

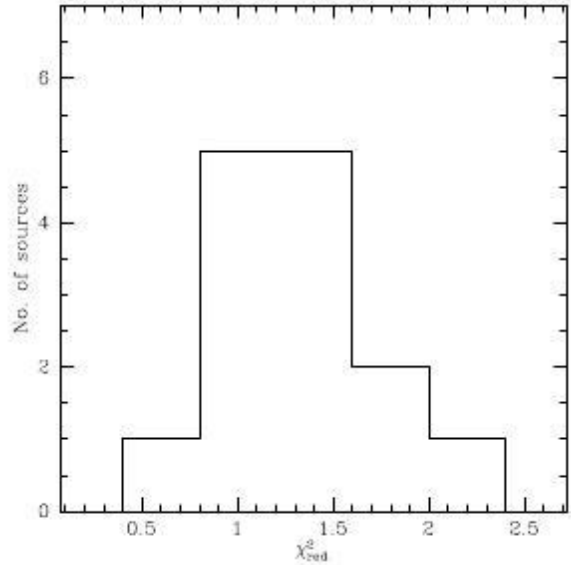

Figure 1: The histogram of reduced $\chi^2$ for sources having soft excess.

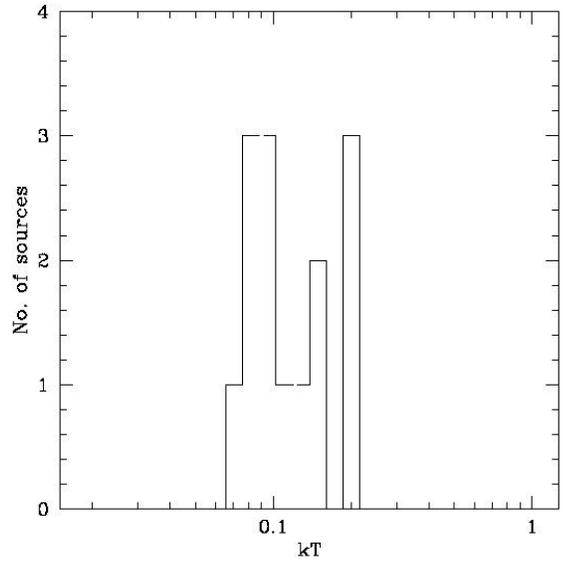

Figure 2: The histogram for the black-body temperature (kT) for sources that have soft excess.

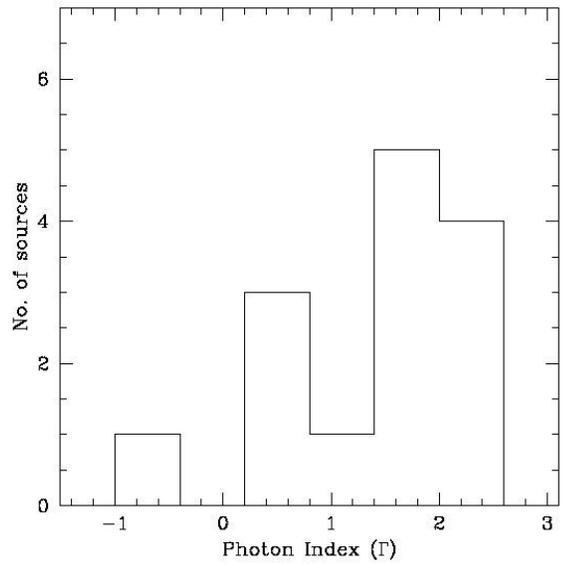

Figure 3: The histogram for the Power-law Photon Index ($\Gamma$) for those sources that have soft excess.

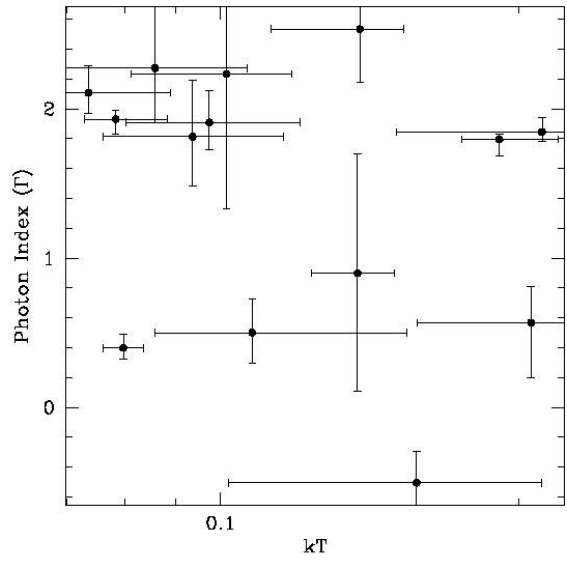

Figure 4: Power-law photon index ($\Gamma$), versus, black-body temperature ($kT$).

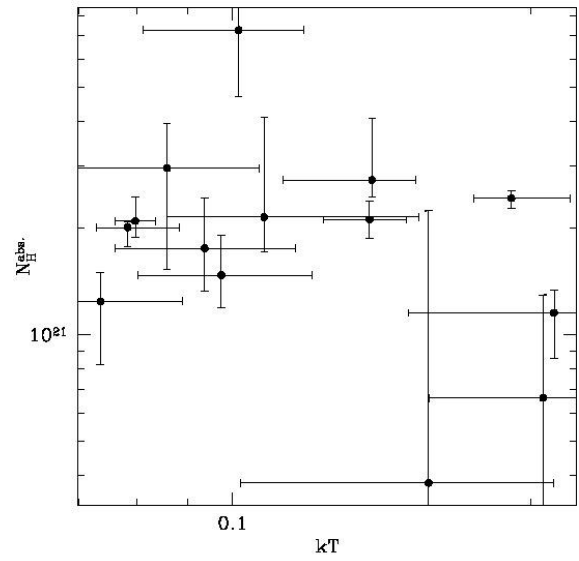

Figure 5: The absorption $N_H^{abs.}$, versus, black-body temperature ($kT$).

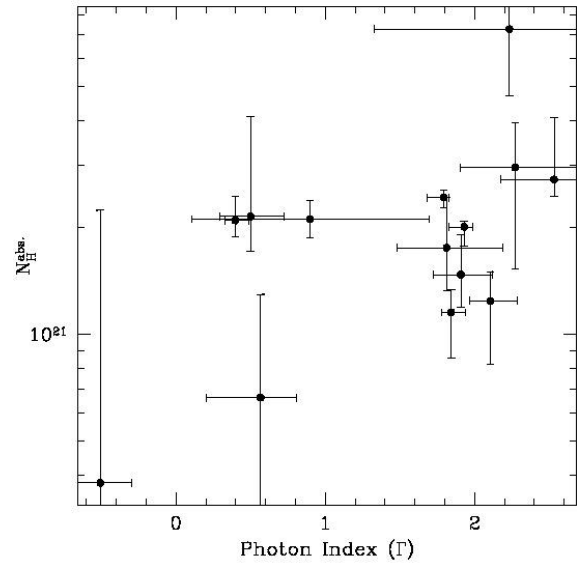

Figure 6: The absorption $N_H^{abs.}$ ,versus, power-law photon index ($\Gamma$).

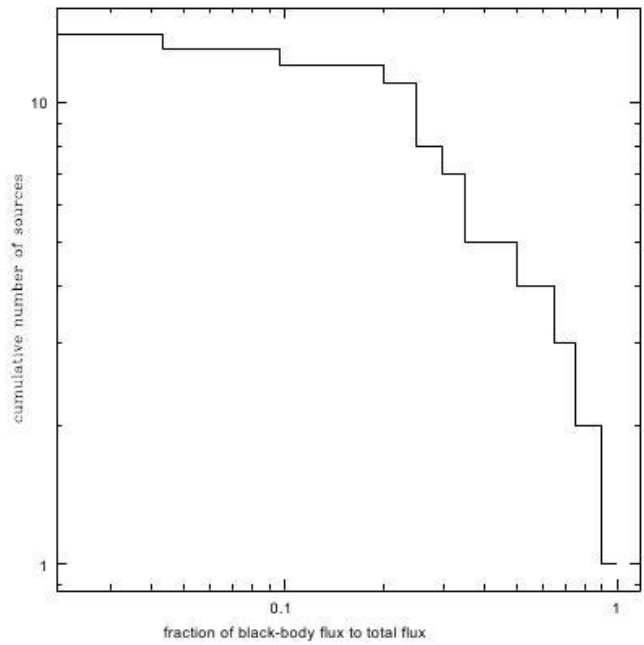

Figure 7: Cumulative histogram for the fraction of black-body flux to the total flux in 0.3-10 keV.

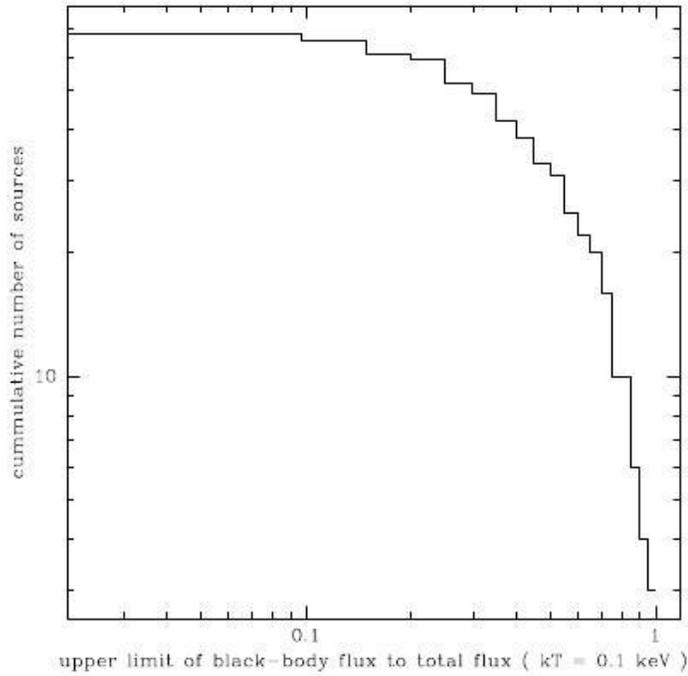

Figure 8: *Cumulative histogram for the upper limit of the fraction of black-body flux to the total flux, in 0.3-10 keV for KT=0.2 keV.*

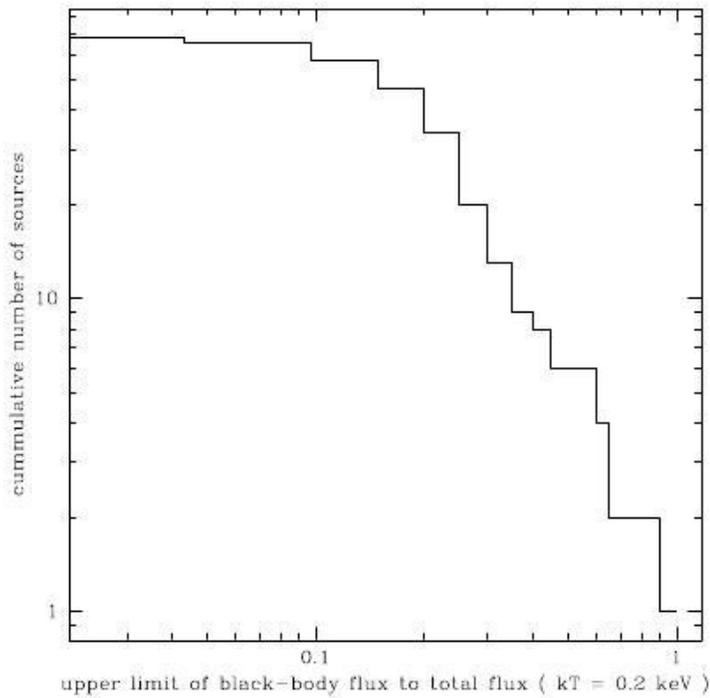

Figure 9: *Cumulative histogram for the upper limit of the fraction of black-body flux to the total flux, in 0.3-10 keV for kT=0.1 keV.*

Of the sample, 67 of the sources do not show evidence for a soft excess, based on the F-test probability at a 99% confidence level. These sources were also fitted with the APLBB model with the black body temperature fixed at 0.1 and 0.2 keV. The normalization of the black body was increased till $\Delta\chi^2 = 2.7$, which provided an upper limit on a soft excess that may be present in these temperatures. The upper limit on the fraction of soft excess flux for these two temperatures are tabulated in Table (4) along with other spectral parameters and NED identification. Figures (7) and (8) show the accumulative histogram for the upper limit of these fractions.

Twenty of the sources are identified by NED as quasars or AGN. The upper limit on the soft excess fraction for these sources is high $> 0.1$, except for one Seyfert 1 AGN where and upper limit of $0.06$ can be imposed. This suggests that many of these sources may indeed have soft excess at a similar fractional level to the sources discussed in the previous section. Thus the statistics do not allow for a good estimate of the fraction of sources that have soft excess.

An interesting object in the present sample is source no. 8 (R.A. 08:32:28.338, DEC +52:36:23.08) which Mrk 0091. This is a ``normal'' galaxy which is known to be X-ray bright (Georgakakis et. al. 2004). Here we see that the spectral index of this galaxy is very steep $\Gamma : 6.5$, and can actually be better fitted ($\chi^2$/d.o.f $= 48.21/28$) with just a black body emission with temperature $kT : 0.183 keV$. The similarity with soft sources discovered in the previous section can lead to the speculation that those are also ``normal'' galaxies that are X-ray bright.

Table 4: X-ray sources that do not have soft excess in the XMM-Newton fields under study, all parameter for APL model.

| $RA_x$[1] (j2000) | $DEC_x$[1] (j2000) | $\chi^2$ / dof | $N_H^{abs.}$ | $\Gamma$[3] | frac.[4] | frac.[5] | separ.[6] arcsec. | Type[7] |
|---|---|---|---|---|---|---|---|---|
| 08:30:27.010 | +52:46:00.68 | 21.8 / 23 | $0.11^{+0.05}_{-0.06}$ | $2.0^{+0.2}_{-0.2}$ | 0.60 | 0.20 | 2.32 | G |
| 08:30:49.829 | +52:49:09.28 | 94.4 / 96 | $0.06^{+0.02}_{-0.06}$ | $1.8^{+0.1}_{-0.1}$ | 0.12 | 0.09 | 1.31 | G |
| 08:31:17.551 | +52:48:55.32 | 36.6 / 46 | $0.04^{+0.03}_{-0.04}$ | $1.7^{+0.2}_{-0.1}$ | 0.40 | 0.15 | 0.82 | G |
| 08:31:41.834 | +52:54:11.98 | 35.1 / 24 | $0.06^{+0.05}_{-0.06}$ | $1.8^{+0.2}_{-0.2}$ | 0.25 | 0.23 | 1.07 | UvES |
| 08:32:05.556 | +52:43:52.93 | 35.2 / 50 | $0.05^{+0.04}_{-0.05}$ | $2.1^{+0.3}_{-0.2}$ | 0.24 | 0.13 | 6.83 | QSO |
| 08:32:18.482 | +52:37:25.48 | 36.1 / 34 | $0.04^{+0.03}_{-0.04}$ | $2.0^{+0.2}_{-0.1}$ | 0.15 | 0.18 | 2.61 | G |
| 08:32:23.283 | +52:48:36.97 | 36.8 / 29 | $0.08^{+0.05}_{-0.08}$ | $1.9^{+0.2}_{-0.2}$ | 0.51 | 0.15 | | |
| 08:32:28.338 | +52:36:23.08 | 50.8 / 28 | $0.70^{+0.23}_{-0.17}$ | $6.3^{+1.6}_{-0.6}$ | 0.90 | 0.61 | 2.42 | G |
| 08:32:28.807 | +52:45:32.74 | 30.7 / 24 | $0.31^{+0.12}_{-0.07}$ | $3.6^{+0.7}_{-0.4}$ | 0.69 | 0.55 | | |
| 08:32:34.355 | +52:34:19.58 | 28.4 / 34 | $0.04^{+0.02}_{-0.04}$ | $1.7^{+0.1}_{-0.1}$ | 0.22 | 0.04 | | |
| 08:32:37.114 | +52:46:10.71 | 44.2 / 27 | $0.08^{+0.06}_{-0.03}$ | $2.5^{+0.6}_{-0.3}$ | 0.71 | 0.34 | | |
| 08:32:38.494 | +52:46:49.33 | 22.8 / 23 | $0.04^{+0.06}_{-0.04}$ | $2.4^{+0.5}_{-0.2}$ | 0.33 | 0.22 | | |
| 09:57:32.078 | +02:43:03.53 | 66.7 / 63 | $0.07^{+0.03}_{-0.03}$ | $1.8^{+0.2}_{-0.2}$ | 0.30 | 0.07 | | |
| 09:58:19.955 | +02:29:05.18 | 46.3 / 46 | $0.03^{+0.03}_{-0.03}$ | $2.1^{+0.2}_{-0.1}$ | 0.14 | 0.09 | 2.33 | QSO |
| 09:58:23.578 | +02:41:07.37 | 28.8 / 26 | $0.03^{+0.05}_{-0.03}$ | $1.6^{+0.3}_{-0.2}$ | 0.25 | 0.07 | | |
| 09:58:31.584 | +02:49:02.71 | 33.8 / 43 | $0.06^{+0.04}_{-0.03}$ | $2.2^{+0.3}_{-0.2}$ | 0.45 | 0.26 | | |
| 09:58:34.032 | +02:44:28.09 | 92.4 / 54 | $0.05^{+0.03}_{-0.05}$ | $2.1^{+0.2}_{-0.2}$ | 0.55 | 0.23 | 1.19 | QSO |
| 09:58:37.385 | +02:36:04.66 | 33.1 / 40 | $0.24^{+0.05}_{-0.06}$ | $1.9^{+0.2}_{-0.2}$ | 0.43 | 0.13 | | |
| 09:58:37.671 | +02:50:34.44 | 33.9 / 27 | $0.11^{+0.05}_{-0.06}$ | $1.9^{+0.2}_{-0.2}$ | 0.71 | 0.18 | | |
| 09:58:48.898 | +02:34:42.75 | 33.0 / 23 | $0.12^{+0.07}_{-0.05}$ | $2.2^{+0.5}_{-0.3}$ | 0.62 | 0.23 | 1.75 | QSO |
| 09:58:52.078 | +02:51:58.15 | 40.0 / 34 | $0.04^{+0.04}_{-0.04}$ | $2.3^{+0.4}_{-0.2}$ | 0.47 | 0.23 | 2.18 | QSO |
| 09:59:08.318 | +02:43:12.33 | 49.6 / 52 | $0.09^{+0.04}_{-0.03}$ | $2.2^{+0.3}_{-0.2}$ | 0.38 | 0.07 | 2.35 | QSO |
| 11:17:35.136 | +07:43:32.73 | 26.2 / 24 | $0.04^{+0.04}_{-0.04}$ | $1.8^{+0.3}_{-0.2}$ | 0.37 | 0.18 | | |
| 11:17:40.296 | +07:44:09.84 | 27.4 / 24 | $0.06^{+0.06}_{-0.06}$ | $2.0^{+0.4}_{-0.3}$ | 0.58 | 0.24 | 26.21 | G |
| 11:17:50.858 | +07:57:09.36 | 149.6 / 140 | $0.12^{+0.02}_{-0.02}$ | $1.8^{+0.1}_{-0.1}$ | 0.06 | 0.10 | 3.53 | G |
| 11:17:59.400 | +07:44:03.66 | 46.0 / 51 | $0.04^{+0.03}_{-0.04}$ | $2.0^{+0.1}_{-0.1}$ | 0.33 | 0.13 | | |
| 11:18:16.337 | +07:43:15.37 | 30.0 / 29 | $0.04^{+0.06}_{-0.04}$ | $1.7^{+0.3}_{-0.2}$ | 0.15 | 0.23 | 9.47 | G |
| 11:18:21.101 | +07:38:14.60 | 25.6 / 29 | $0.11^{+0.06}_{-0.05}$ | $2.4^{+0.4}_{-0.3}$ | 0.74 | 0.35 | | |
| 11:18:22.344 | +07:44:48.51 | 41.0 / 46 | $0.06^{+0.03}_{-0.03}$ | $1.9^{+0.1}_{-0.1}$ | 0.50 | 0.20 | 3.66 | G |
| 11:18:38.218 | +07:42:41.73 | 26.7 / 30 | $0.04^{+0.04}_{-0.04}$ | $1.6^{+0.2}_{-0.1}$ | 0.35 | 0.12 | | |
| 11:19:10.855 | +07:39:11.76 | 15.6 / 22 | $0.04^{+0.05}_{-0.04}$ | $1.6^{+0.3}_{-0.2}$ | 0.54 | 0.14 | | |
| 12:02:14.456 | -21:01:55.69 | 33.3 / 35 | $0.19^{+0.11}_{-0.08}$ | $2.6^{+0.7}_{-0.4}$ | 0.80 | 0.41 | | |
| 12:02:20.815 | -20:52:19.66 | 65.5 / 45 | $0.20^{+0.08}_{-0.05}$ | $3.0^{+0.6}_{-0.3}$ | 0.73 | 0.25 | 0.40 | RadioS |
| 12:02:24.361 | -20:51:59.81 | 33.0 / 22 | $0.29^{+0.13}_{-0.14}$ | $1.8^{+0.6}_{-0.3}$ | 0.83 | 0.14 | | |
| 12:02:35.395 | -20:57:38.52 | 20.2 / 22 | $1.8^{+3.6}_{-1.3}$ | $4.5^{+1.1}_{-2.5}$ | 1.00 | 1.00 | | |
| 12:02:40.973 | -20:49:53.54 | 26.3 / 25 | $0.06^{+0.06}_{-0.04}$ | $2.3^{+0.5}_{-0.3}$ | 0.53 | 0.16 | | |
| 12:02:46.472 | -21:02:26.15 | 23.4 / 30 | $0.25^{+0.18}_{-0.17}$ | $2.1^{+1.1}_{-0.5}$ | 0.84 | 0.31 | | |
| 12:02:48.609 | -21:06:19.36 | 41.3 / 47 | $0.07^{+0.06}_{-0.07}$ | $2.2^{+0.4}_{-0.3}$ | 0.21 | 0.17 | | |
| 12:02:52.030 | -20:53:11.80 | 11.4 / 22 | $0.04^{+0.06}_{-0.04}$ | $2.0^{+0.4}_{-0.2}$ | 0.31 | 0.26 | | |
| 12:03:05.217 | -20:54:42.39 | 17.9 / 30 | $0.04^{+0.08}_{-0.04}$ | $2.4^{+0.6}_{-0.4}$ | 0.67 | 0.28 | | |
| 12:03:09.282 | -20:46:19.85 | 33.4 / 35 | $0.04^{+0.06}_{-0.04}$ | $1.8^{+0.3}_{-0.2}$ | 0.73 | 0.23 | | |
| 12:03:13.965 | -20:50:54.78 | 21.4 / 24 | $0.13^{+0.45}_{-0.13}$ | $2.3^{+2.4}_{-0.9}$ | 0.97 | 0.56 | | |
| 12:03:16.036 | -20:54:04.72 | 22.1 / 35 | $0.04^{+0.08}_{-0.04}$ | $1.9^{+0.6}_{-0.3}$ | 0.66 | 0.22 | | |
| 12:03:17.569 | -20:58:57.41 | 46.0 / 33 | $0.04^{+0.10}_{-0.04}$ | $2.8^{+1.0}_{-0.3}$ | 0.39 | 0.44 | | |
| 12:03:25.754 | -20:55:03.66 | 36.6 / 26 | $1.7^{+1.8}_{-1.0}$ | $2.2^{+1.5}_{-0.5}$ | 0.99 | 0.89 | | |
| 12:03:38.999 | -20:54:07.63 | 51.3 / 35 | $0.04^{+0.02}_{-0.04}$ | $1.9^{+0.2}_{-0.2}$ | 0.58 | 0.18 | | |
| 12:03:44.895 | -21:03:25.81 | 17.9 / 28 | $0.04^{+0.09}_{-0.04}$ | $2.2^{+0.6}_{-0.3}$ | 0.39 | 0.28 | | |
| 12:03:45.382 | -20:50:47.41 | 18.9 / 27 | $0.10^{+0.11}_{-0.10}$ | $1.9^{+0.9}_{-0.5}$ | 0.37 | 0.27 | | |
| 21:14:40.185 | +06:00:15.83 | 20.8 / 23 | $0.27^{+0.15}_{-0.10}$ | $3.1^{+1.8}_{-0.8}$ | 0.90 | 0.63 | | |
| 21:14:54.525 | +06:16:55.38 | 24.3 / 25 | $0.07^{+0.07}_{-0.07}$ | $1.4^{+0.2}_{-0.2}$ | 0.73 | 0.20 | 1.97 | RadioS |

| 1 | 2 | 3 | | | 4 | 5 | 6 | 7 |
|---|---|---|---|---|---|---|---|---|
| 21:14:59.395 | +06:06:29.75 | 59.3 / 66 | $0.26^{+0.04}_{-0.04}$ | $2.0^{+0.1}_{-0.1}$ | 0.44 | 0.21 | 9.79 | XrayS |
| 21:15:07.913 | +06:07:22.22 | 45.5 / 39 | $0.12^{+0.06}_{-0.12}$ | $2.0^{+0.2}_{-0.2}$ | 0.81 | 0.34 | 16.23 | RadioS |
| 21:15:18.849 | +06:06:46.92 | 27.5 / 25 | $0.07^{+0.05}_{-0.07}$ | $1.2^{+0.2}_{-0.2}$ | 0.41 | 0.16 | 1.19 | QSO |
| 22:14:53.113 | -17:42:35.35 | 33.1 / 29 | $0.06^{+0.05}_{-0.04}$ | $2.2^{+0.4}_{-0.3}$ | 0.32 | 0.19 | 2.85 | XrayS |
| 22:14:56.679 | -17:50:53.42 | 44.1 / 39 | $0.46^{+0.18}_{-0.11}$ | $2.0^{+0.2}_{-0.3}$ | 0.93 | 0.17 | 0.65 | XrayS |
| 22:15:15.114 | -17:32:24.54 | 136.3 / 109 | $0.02^{+0.01}_{-0.02}$ | $2.5^{+0.1}_{-0.1}$ | 0.24 | 0.02 | 5.67 | QSO |
| 22:15:19.451 | -17:51:12.65 | 49.4 / 34 | $0.07^{+0.04}_{-0.03}$ | $2.2^{+0.4}_{-0.2}$ | 0.51 | 0.24 | 10.38 | XrayS |
| 22:15:23.580 | -17:43:20.76 | 83.7 / 69 | $0.05^{+0.03}_{-0.02}$ | $2.2^{+0.2}_{-0.1}$ | 0.30 | 0.18 | 2.82 | G |
| 22:15:36.622 | -17:33:56.48 | 28.2 / 26 | $0.06^{+0.07}_{-0.06}$ | $2.3^{+0.5}_{-0.4}$ | 0.24 | 0.29 | 1.82 | VisS |
| 22:15:38.110 | -17:46:33.22 | 47.4 / 48 | $0.02^{+0.03}_{-0.02}$ | $2.0^{+0.2}_{-0.1}$ | 0.22 | 0.07 | 17.43 | G |
| 22:15:50.326 | -17:52:08.93 | 55.1 / 63 | $0.02^{+0.02}_{-0.02}$ | $2.1^{+0.1}_{-0.1}$ | 0.10 | 0.08 | 1.06 | VisS |
| 22:16:02.014 | -17:39:48.16 | 36.9 / 33 | $0.07^{+0.06}_{-0.07}$ | $1.9^{+0.3}_{-0.4}$ | 0.65 | 0.32 | | |
| 22:16:03.071 | -17:43:18.49 | 29.4 / 39 | $0.05^{+0.05}_{-0.05}$ | $2.0^{+0.4}_{-0.3}$ | 0.56 | 0.26 | 5.11 | XrayS |
| 22:16:04.657 | -17:52:19.41 | 19.6 / 28 | $0.04^{+0.06}_{-0.04}$ | $2.2^{+0.4}_{-0.3}$ | 0.17 | 0.15 | 4.22 | VisS |
| 22:16:09.408 | -17:46:42.11 | 25.1 / 25 | $0.08^{+0.07}_{-0.03}$ | $2.9^{+0.6}_{-0.3}$ | 0.54 | 0.14 | | |
| 22:16:23.436 | -17:43:17.20 | 67.0 / 57 | $0.10^{+0.04}_{-0.04}$ | $2.2^{+0.3}_{-0.2}$ | 0.35 | 0.13 | 0.94 | G |
| 22:16:23.488 | -17:47:24.36 | 48.5 / 37 | $0.03^{+0.05}_{-0.03}$ | $2.1^{+0.4}_{-0.2}$ | 0.11 | 0.16 | 3.09 | VisS |

1-position of the X-ray source obtained by *edetect_chain,* 2-APL model. 3-Power-law photon index, 4-upper limit of the fraction of black-body flux to total flux for *kT* = 0.1 keV, 5-upper limit of the fraction of black-body flux to total flux for *kT* = 0.2 keV, 6- separation of X-ray source position from the NED position in arcsec., 7-Type of the best matched sources with our sample from NED

## 3 Conclusion

We selected six XMM-Newton fields such that the observations had an exposure time freed of flares (i.e. GTI ) greater than 60 ksec, had data from the EPIC-pn detector operating in full frame mode and were for high galactic latitude $|b| > 25^o$. Bright sources (with at least $500$ net counts in the $0.3 - 10$ keV band) were selected from these fields for spectral analysis, only four bright sources have been rejected because they are diffuse. The emphasis of the analysis is to study the soft excess in AGN spectra. We estimate the fraction of sources that exhibit soft excess at a 99% confidence level. This has been done by applying absorbed power-law model and then an absorbed power-law with a black body model to all spectra and the F-test probability for the additional component was tested. Those sources for which the F-test probability was less than 0.01 (i.e. at a confidence level of 99%) were considered to have a soft excess. The 81 sources of the sample were cross-correlated with the NED catalogue for identification.

Our results show that 14 of the 81 sources in the sample display evidence of soft excess. We examine for correlations between the soft excess temperature, photon index and absorption column density and find visually weak possible indications. $N_H^{abs.}$ was found to correlate with $\Gamma$ which may be attributed to the artifacts of low count statistics. Moreover, statistical analysis of these correlation reveal that the results are dominated by a few bright sources biasing the sample and it was difficult to draw definite conclusions. The observed soft excess temperature spread for these sources is narrow, ranging only from $0.08$ to $0.2$ keV. Six of these sources were identified in NED as AGN/QSO with red-shifts ranging from $0.022$ to $3.9$. Thus the intrinsic spread of soft excess temperature ranges from $0.08-1.0$ keV. Three of the fourteen sources have very soft spectra, with the black body spectrum contributing more than 70% of the flux. These soft sources have high inferred column density $>2\times10^{21}\ cm^{-2}$. For the rest of the sources, the soft excess fraction ranges from $0.03-0.5$.

Sources, which did not satisfy the 99% confidence level criterion for the existence of a soft excess, upper limits on the fractional soft excess flux were obtained assuming that the soft excess temperature was $0.1$ and $0.2$ keV. In general, the upper limit turns out to be $>0.1$ keV**,** and hence it is possible that these sources also have soft excess similar to the ones where the excess was detected. One of the sources for which soft excess was not detected was Mrk 0091. This is a known 'normal' galaxy which is bright in X-rays. However, the spectral index of the power-law for this source is large, $\Gamma:6.5$, and it can be fitted with a black body spectrum with $kT:0.183 keV$. The similarity with other soft but unidentified sources detected in this sample suggests that these objects belong to the same class.

The sample under study needs to be extended to deal with more objects, including more quasars, more AGN. Time variability of the soft excess can shed light on the mechanism producing such variability. Matching with other catalogues and archives (e.g. SIMBAD, USNO, APM-North,…etc.) will be done in the future taking into account the ratio of X-ray flux ($f_x$) to optical flux ($f_{op}$) in order to improve the reliability of the identifications. The specifically interesting sources with detectable soft sources and those whose total spectrum can be described as soft thermal emission, need to be followed up with multi-wavelength analysis.

## APPENDIX A

These Figures show the digitized images of our fields overlayed by NED catalogue (in green), our sources (in red), matched sources (in blue), and the EPIC-pn instrument FOV.

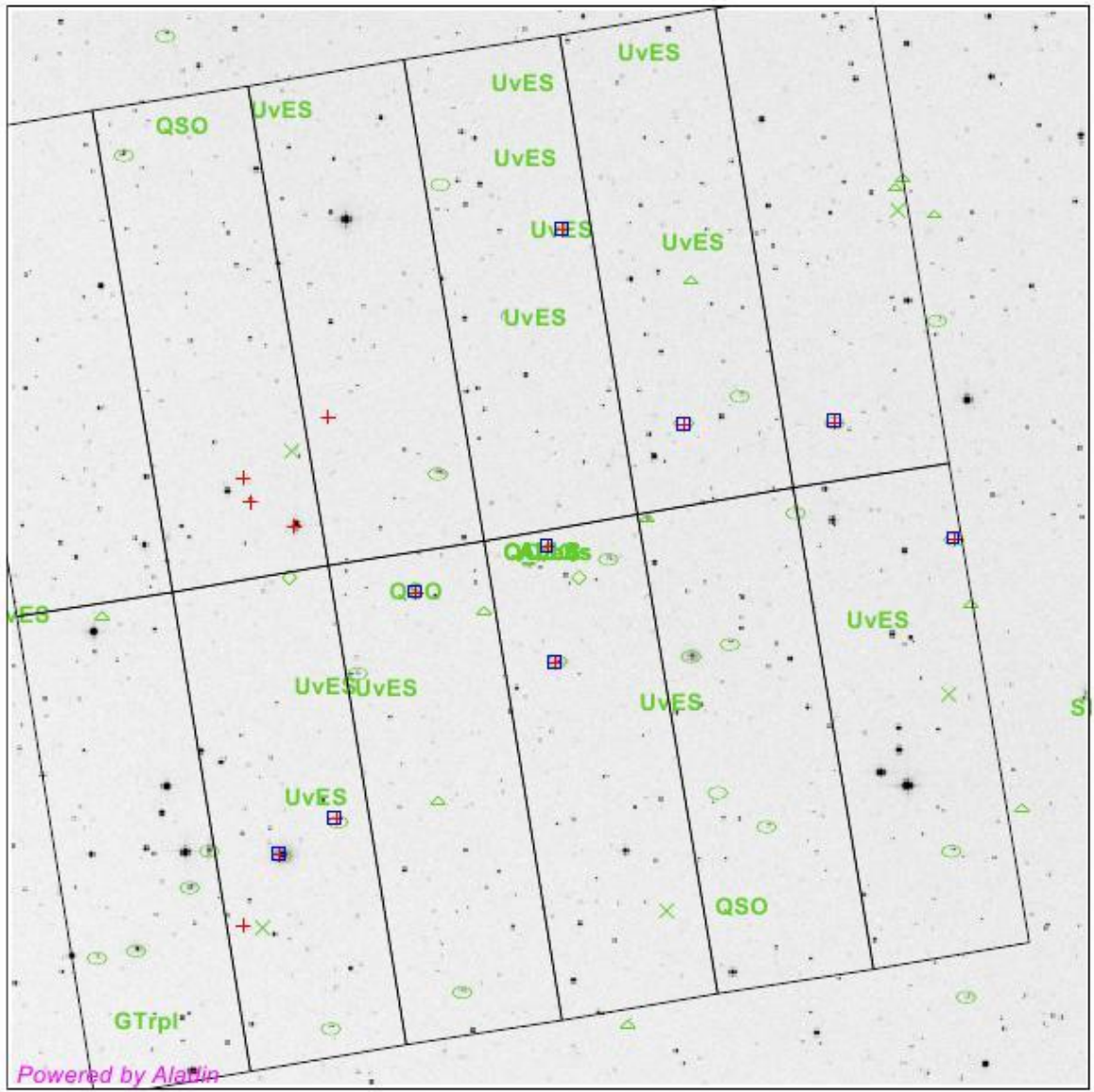

Figure 1: *Matching results for sources in APM 28279+5255 field: NED (green), our sources (red), matched sources (blue), and EPIC-pn FOV*

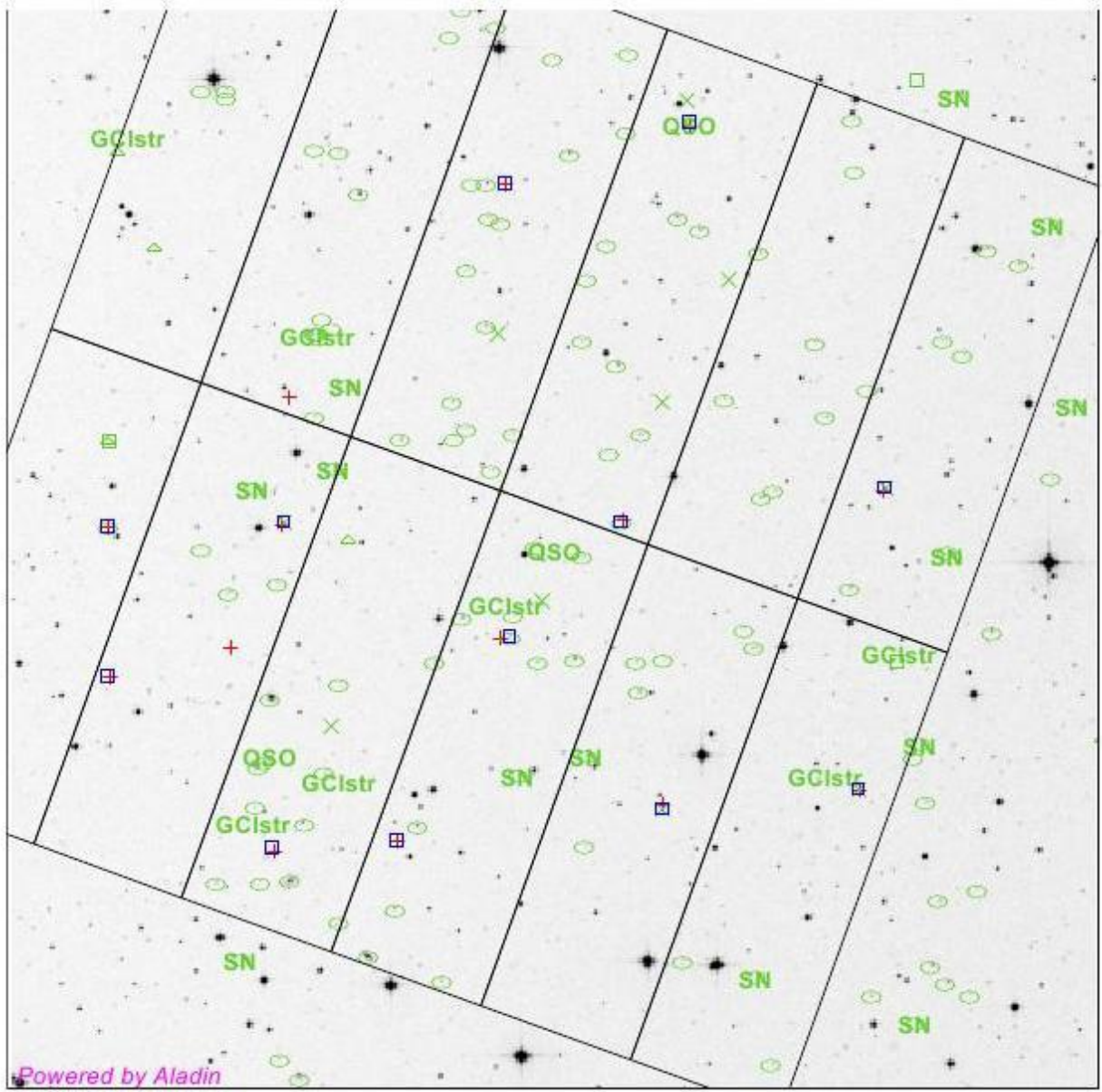

Figure 2: *Matching results for sources in LBQS 2212-1759 field: NED (green), our sources (red), matched sources (blue), and EPIC-pn FOV*

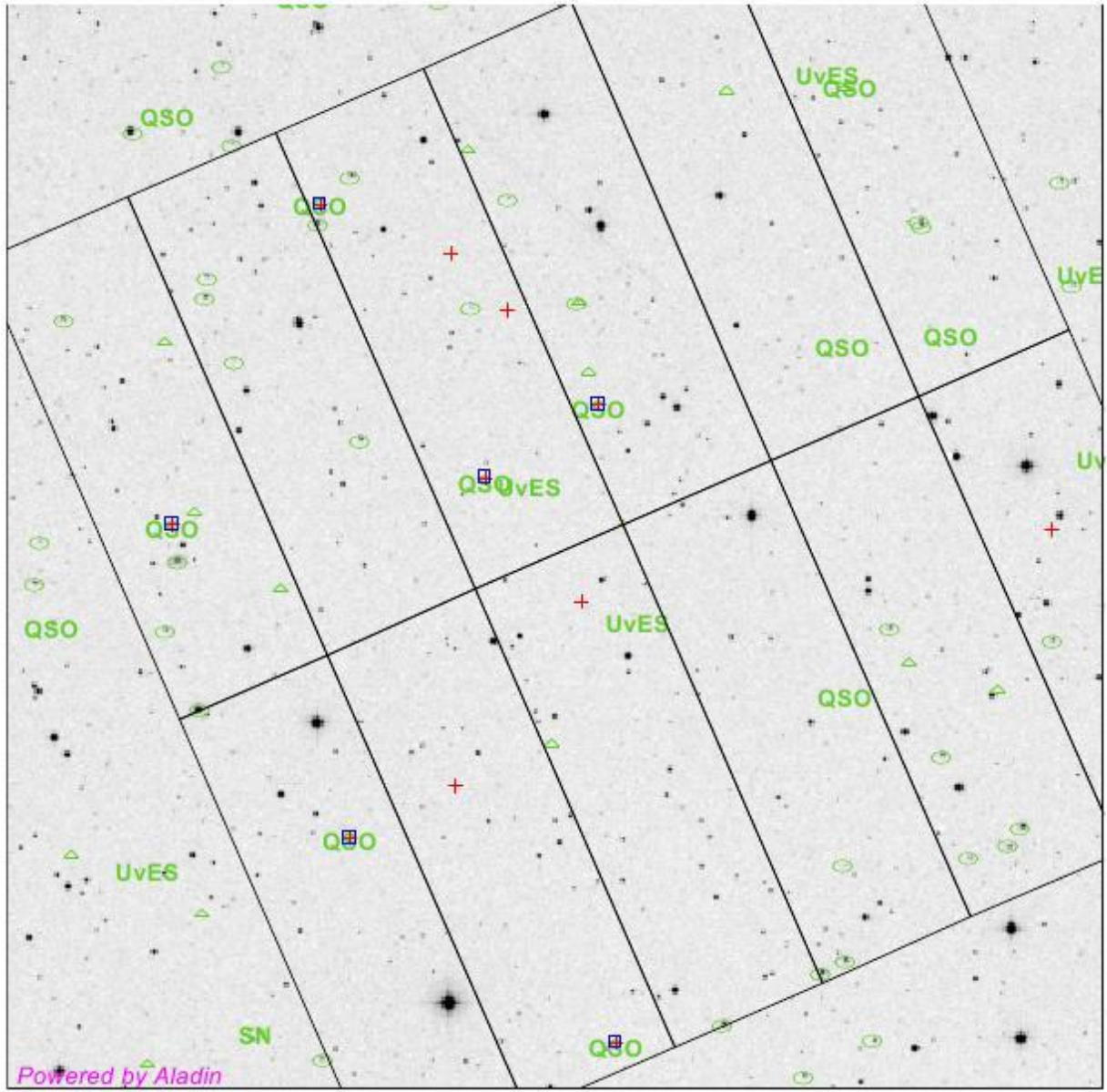

Figure 3: *Matching results for sources in COSMOS FIELD 21 field: NED (green), our sources (red), matched sources (blue), and EPIC-pn FOV*

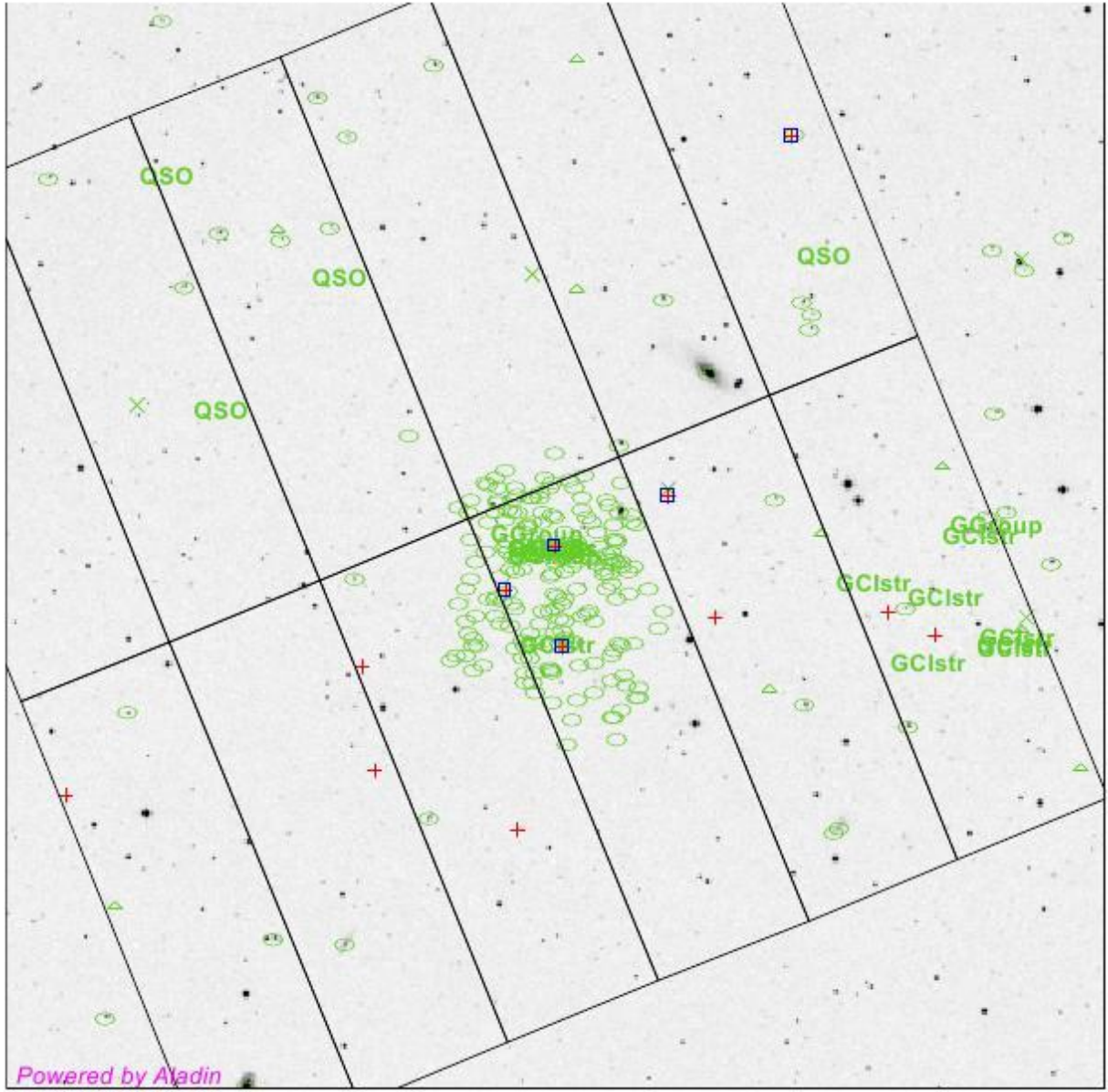

Figure 4: *Matching results for sources in PG 1115+080 field: NED (green), our sources (red), matched sources (blue), and EPIC-pn FOV*

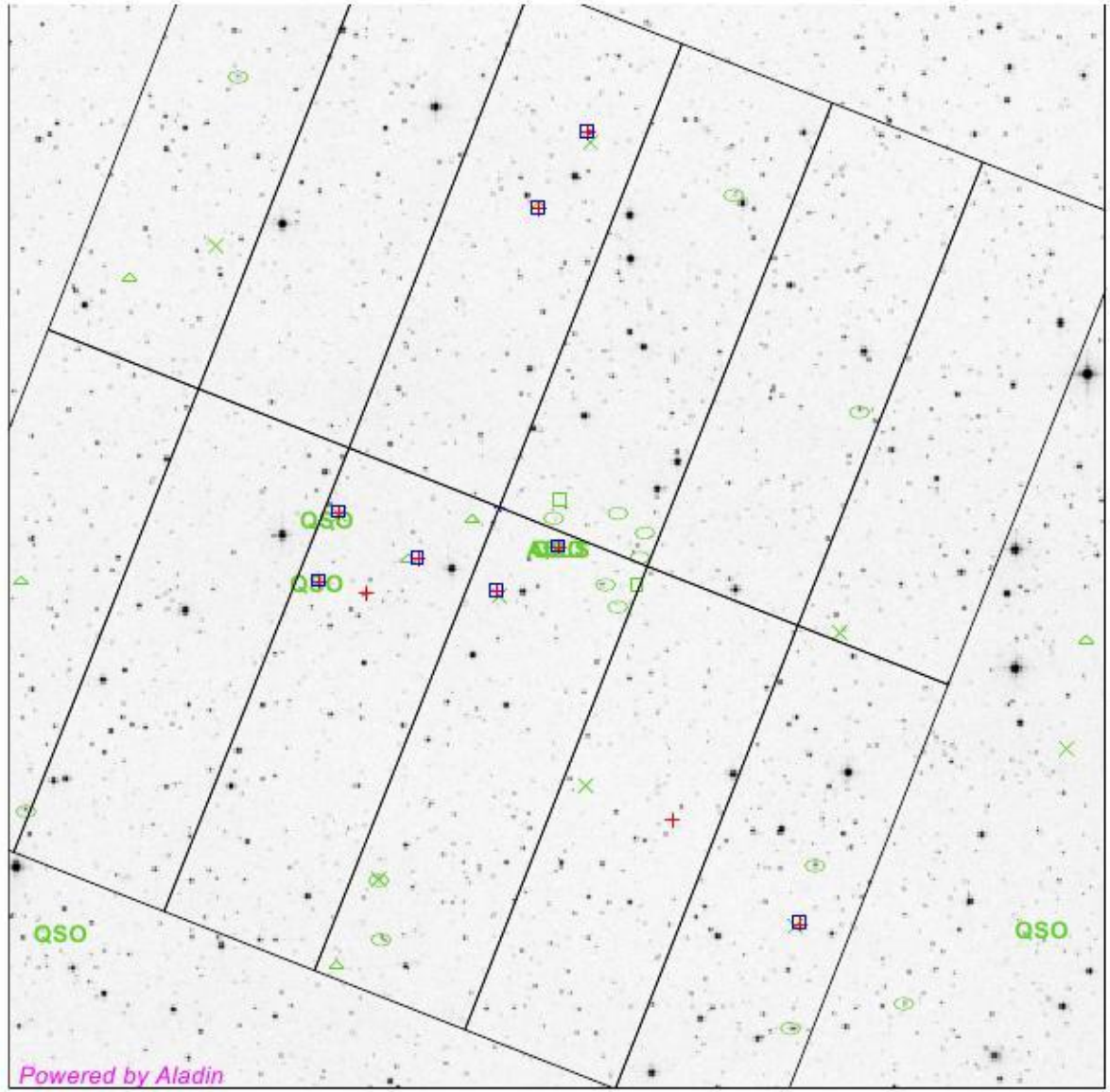

Figure 5: *Matching results for sources in PG 2112+059 field: NED (green), our sources (red), matched sources (blue), and EPIC-pn FOV*

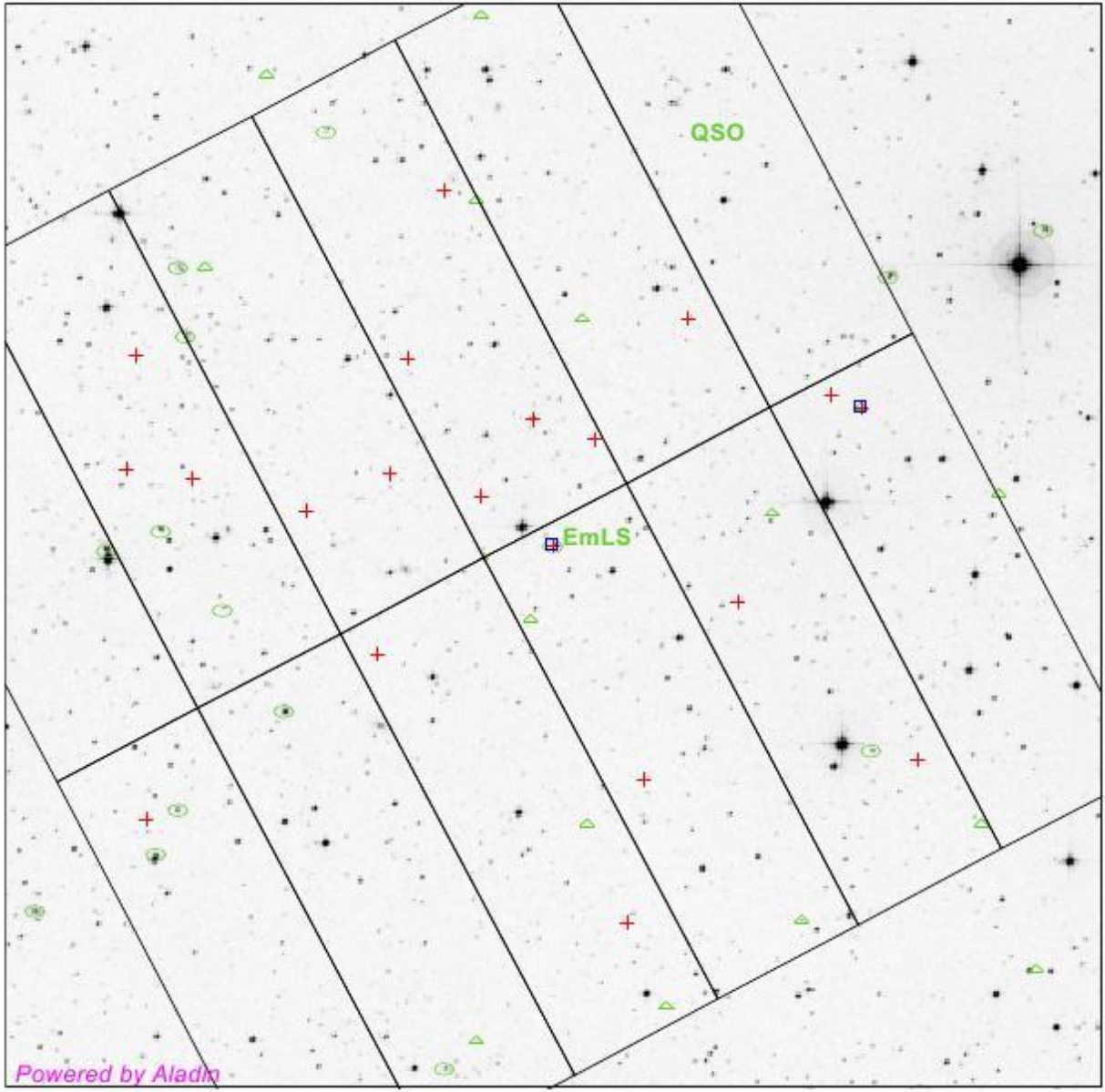

Figure 6: *Matching results for sources in POX52 field: NED (green), our sources (red), matched sources (blue), and EPIC-pn FOV*